# Capacitance investigation of Ge nanoclusters on a silicon (001) surface grown by MBE at low temperatures

O.V. Feklisova [a,*], E.B. Yakimov [a], L.V. Arapkina [b], V.A. Chapnin [b], K.V. Chizh [b], V.P. Kalinushkin [b], V.A. Yuryev [b]

[a] *Institute of Microelectronics Technology RAS, Chernogolovka, 142432, Russia*

[b] *A.M. Prokhorov General Physics Institute RAS, Moscow, 119991, Russia*

**Abstract**

Electrical properties of multilayer arrays of germanium nanoclusters grown on the silicon (001) surface at low temperature have been studied. A correlation between the quantum dot (QD) density estimated from STM and the charge accumulated by QD layers as extracted from *C-V* characteristics was revealed. Temperature dependence of the *C-V* characteristics was studied. At low temperatures, the hysteresis was observed. DLTS spectra measured on the structures with different QD arrangements demonstrated essential distinctions.

## 1. Introduction

Tempting possibilities of the self-assembled quantum dots for optoelectronic device fabrication stimulate extensive experimental and theoretical investigations of these objects. In the recent years significant progress was achieved in structural quality and high homogeneity of self-assembled quantum dots grown by the molecular-beam epitaxy (MBE) method. The density, sizes and morphology of QDs have been shown to be governed by the growth conditions [1]. Relatively low growth temperatures (350–500°C) allow forming three-dimensional islands with nanometer sizes and higher density (up to $10^{12}$ cm$^{-2}$). These parameters are determinant for the electronic structure of QDs and their arrays. Thus, deep levels assosiated with QDs and their capture and emission characterictics are very important. Electrical spectroscopy techniques such as capacitance-voltage (C-V), admittance and deep level transient spectroscopy (DLTS) may be applied in order to investigate these parameters of QDs [2-5].

In the present work, the electrical properties of Ge nanocluster arrays grown at low temperatures with different germanium coverages was ivestigated by capacitance-voltage and deep level transient spectroscopy techniques. Results obtained were compared with the QD structure studied by scanning tunneling microscopy (STM).

## 2. Experimental

A series of samples with Ge-rich dots was grown on atomically clean B-doped Cz Si (100) wafers ($\rho$ = 12 $\Omega$ cm) in integrated ultra-high-vacuum (UHV) system based on the Riber EVA 32 molecular beam epitaxy chamber coupled with the GPI-300 scanning tunnelling microscope [6]. This instrument enables the STM investigation of samples at any stage of Si surface cleaning and MBE growth without leaving UHV ambient. On a 100-nm thick Si buffer, five layers of different Ge coverage ($h_{Ge}$ = 6, 10 or 14 Å) were embedded separated by 50 nm of nominally undoped Si layers and covered with a 100-mn thick Si cap. The samples will be referred to as QD6, QD10 and QD14 in accordance with the value of $h_{Ge}$.

The deposition temperature was 350°C for Ge layers and 530°C for Si spacer layers. The lateral dimensions, heights and surface densities of QDs were determined by STM on the satellite samples.

Al Schottky diodes were deposited on the top of mesa etched structures for capacitance (*C*) and conductance (*G*) measurements. The characteristics were measured at

---

* Corresponding author. Tel.: +7-496-524-4091; fax: +7-495-962-8047; e-mail: feklisov@iptm.ru

1 MHz test signal frequency with 15-mV amplitude using an EG&G C-V Plotter model 410 with respect to either the capacitive current component $C(V)$ or the conductance current component $G(V)$. Retrace operation (from reverse to forward bias and back) facilitates hysteresis effect studies. The standard DLTS setup including a lock-in amplifier as the correlator was used for deep level measurements.

## 3. Results and discussion

Figure 1 (a, b, c) shows the STM images of the Ge/Si(001) heterostructures with different $h_{Ge}$. One can see that QD's morphology, size and density change depending on the effective thickness of Ge layer. It is seen that at $h_{Ge}$ = 6 Å (QD6) the Ge "hut" clusters are formed and have pyramidal or wedge shape, the latter dominating. Typical QD's sizes are as follows: height ($h$) is 0.6÷1 nm and width ($l$) is 7÷8 nm; the cluster lengths are spread rather uniformly over a wide interval of values. The Ge wetting layer between the QDs has the developed ($M\times N$) surface (see Fig. 1(a)). The density of QDs is about $3\times 10^{11}$ cm$^{-2}$ for this $h_{Ge}$.

For $h_{Ge}$ = 10 Å (QD10), a dense array of the QDs is formed with the concentration of ~ $6\times 10^{11}$ cm$^{-2}$ and the following sizes of QDs: $h$ ~ 1÷1.5 nm and $l$ ~ 10÷15 nm (Fig. 1(b)). The dot size variation and the wetting layer fraction between QDs are decreased. The pyramids have virtually disappeared. At the same time, QDs with smaller sizes grow, and their heights have reached several monolayers.

The STM image of the QD array formed at $h_{Ge}$ = 14 Å (QD14) is shown on Fig. 1(c). One can see three types of the dots: those with the pyramidal or wedge shape or a shape resembling wedge with two ridges, whose height generally does not exceed 1.5 nm. QD's sizes are close to those in the QD10. The growing successive incomplete facets are seen on the trapezoid faces of the clusters. The wetting layer between the dots is not seen as a new phase of small dots have arisen between the "big" ones. The total density of QDs is around $2\times 10^{11}$ cm$^{-2}$.

So, the dot density increases with $h_{Ge}$, reaches its peak at $h_{Ge}$ = 10 Å and decrease for $h_{Ge}$ = 14 Å. It may be connected with coalescence process when "big" dots absorb "small" ones.

Figure 2 shows *C-V* traces of the samples measured at room temperature. In all structures, the capacitance changes do not follow the $1/V^{1/2}$ law and the plateaus in the *C-V* traces are revealed for each sample. In the QD6 (Fig. 2 (curve 1)) the plateau is observed for bias voltage between 0.7 and 4.5 V. In the QD10 and QD14 (Fig. 2 (curve 2 and 3)), the plateaus are occurred between – 1 and 6 V, –0.7 and 0.5 V, respectively. In addition, no hysteresis effect was revealed in retrace mode. This fact indicates that at room temperature hole capture by the QDs and their emission follow the voltage scan of the capacitance meter.

In Fig. 2 one can see that the plateau of quasi-constant capacitance is different for each sample. From the width $\Delta U$ and $C_p$ value of the quasi-constant capacitance, one can estimate the hole concentration in QDs: $p = \Delta U C_p / Sq$, where $S$ is the Schottky diode area and $q$ is the elementary charge. Using *C-V* data for each structure, the following values of the concentration were obtained: $p \approx 3.4\times 10^{11}$ cm$^{-2}$ for the QD6, $p \approx 7\times 10^{11}$ cm$^{-2}$ for the QD10 and $p \approx 1.7\times 10^{11}$ cm$^{-2}$ for the QD14. These values are in agreement with the dot concentration estimated from STM data, and the amount of holes accumulated in the QDs is proportionate to the QD density.

*C-V* characteristics recorded at different temperatures from 77 K to 300 K do not show any essential features in the QD6 and QD10, whereas in the QD14 the hysteresis in the *C-V* trace (see inset Fig. 2) has been observed in the temperature range from 77 to 200 K. On inset of Fig. 2 one can see that the plateau realized at voltage scanning from reverse to forward bias (capture process) contains two shoulders, which could be associated with the carrier capture into two types of the dots with two different energy levels in this structure. At retrace scanning carrier emission occurs from one of the levels, and only one shoulder can be observed in *C-V* curve. It could be explained that the emission rate from one of the levels (the deepest one) becomes lower, and carriers freeze onto this QD's level.

Using the relations $p = (q\varepsilon_{Si}\varepsilon_0)^{-1} \times C^3/(dC/dV)$ and $W = \varepsilon_{Si}\varepsilon_0 S C^{-1}$, the carrier concentration profiles can be calculated from the *C-V* traces. The apparent carrier distributions of the structures under investigation are shown in Figure 3. From Fig. 3 one can see that for the QD6 structure, the position of concentration peak caused by accumulation of holes and the presence of the corresponding plateau on the *C-V* characteristic is close to the conjectural geometrical position of the deepest QD's layer and is of about 320 nm. Close value has also been obtained for the QD10 ($\approx$ 290 nm). For the QD14 a depth of the concentration peak does not coincide with expected geometrical position of QD's layer and is of about 520 nm. Resembling situation has been observed under capacitance-voltage profiling at room temperature in [7], when thermal emission rate of carriers from the deep levels was quite high; however, there was not enough time to establish quasi-equilibrium at high frequency of measurement. As a result, the concentration peak was shifted relative to the geometric position of the layer toward the edge of the depletion region. Another possible explanation might be the influence of $\lambda$–layer for the traps with the deep levels non-uniformly located in the sample depth, when instead of expected peak in a concentration profile on a certain depth $x$, this peak appears on the depth $(x + \lambda)$ [8]. Using the relation for the $\lambda$–layer one could estimate the trap energy under an assumption that observed deepening ($\approx$ 520 nm) from expected position ($\approx$ 300 nm) is determined by these traps whose $\lambda$ is of about 220 nm. Received value is of about 0.28 eV above the top of the Si valence band, i.e. the presence of such traps could provide the observed deviation from the real geometrical location of the QDs.

In Fig. 4 DLTS spectra measured in the structures under study are shown. In all investigated structures peaks are found by using appropriate reverse bias sequences in DLTS measurements. Values of the reverse bias have been chosen in ranges of the voltage where the plateaus in the *C-V* traces are revealed for each structure. So in DLTS measurements the reverse bias does not exceed 5 V, 6 V and 2 V for for the QD6, QD10 and QD14, respectively. Rather limited conditions for measurements of DLTS spectra in these structures are defined by the rather small doping level of substrates and by the relatively small depth of QD's location. If $V_{rev}$ surpassed appointed value the DLTS signal decreased drastically. Such behavior of the DLTS signal is characteristic for the defects non-uniformly distributed in volume of the sample. Therefore we could assume that the peaks observed in spectra are caused, most likely, by the presence of Ge nanoclusters. Besides, the amplitude of a signal from the defects located in very narrow layer depends on width of the depletion region ($W$) (or reverse bias $V_{rev}$) as $\Delta C \sim W^{-3}$ or $\sim (V_b+V_{rev})^{-3/2}$ where ($V_b+V_{rev}$) is the sum of the built-in and reverse bias. Such dependence was observed for quantum dots in [9] and it was fixed for peak at 235 K in the QD14.

In Fig. 4 (curves 1 and 2) one can see that spectra of the QD6 and QD10 have complicated shapes and the peaks are not resolved. For this reason from frequency-temperature dependence for these peaks the activation energies of the centers with deep levels could not been determined correctly. And only in the QD14 the activation energy ~ 260 meV has been obtained for the peak at 235 K. This value correlates with our estimation for the deep level (~ 0.28 eV), which could provide a deviation from the real geometrical location of the QDs observed in the concentration profile in the QD14. Let us notice that in all structures the peak was detected also at temperature ~ 168K (Fig. 4), which was observed under various conditions of measurements and was caused, most likely, by the presence of bulk defects in the samples.

In all investigated structures the centers with deep levels are revealed and associated, most likely, with the presence of Ge nanoclusters with the exception of the peak at ~ 168K. However, the centers observed in the structures with different Ge thickness have essential differences, which could be defined by distinction in properties of the quantum dots. It is impossible to exclude also a probability of defect formation on Ge/Si interfaces, which could form due to a strain relaxation at QD's growth process. These defects will be located in a narrow layer and may bring forth deep levels in the band gap.

In summary, Ge nanoclusters grown on Si at low temperature with various germanium coverages were investigated by C-V and DLTS methods. The plateaus of quasi-constant capacitance in the C-V characteristics are observed and the estimation of a number of holes accumulated in the QDs is carried out. It has been found that the hole concentrations in the QDs are different for each structure and correlate with the QD density estimated from STM. At low temperatures, the hysteresis was observed in the C-V characteristics of the structure with thick Ge layers. In all investigated structures the centers with the deep levels, located in near surface layer and probably caused by presence of the Ge nanoclusters are revealed.

**Acknowledgement**

The work carried out in IMT RAS was partially supported by the Program of basic researches of RAS No 27 "The fundamentals of basic researches of nanotechnologies and nanomaterials".

**Figure captions**

Fig. 1. STM images of the Ge/Si(001) QD arrays, $h_{Ge}$ = 6 (a), 10 (b) and 14 (c) Å.

Fig. 2. *C-V* characteristics of the structures QD6 (1), QD10 (2) and QD14 (3) measured at room temperature. Insert: *C-V* characteristic of the QD14 measured at 77 K in retrace mode.

Fig. 3. Concentration profiles of QD6 (1), QD10 (2) and QD14 (3) extracted from *C-V* characteristics.

Fig. 4. DLTS spectra measured with rate window 49 s$^{-1}$, pulse duration 1 ms, $V_{rev}/V_{pulse}$ = 3 V/–3 V for QD6 (1), $V_{rev}/V_{pulse}$ = 5 V/–5 V for QD10 (2) and $V_{rev}/V_{pulse}$ = 2 V/–2.5 V QD14 (3).

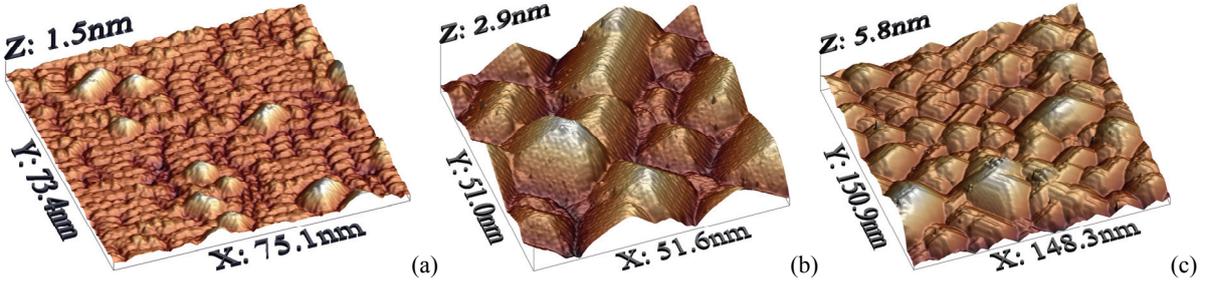

Figure 1

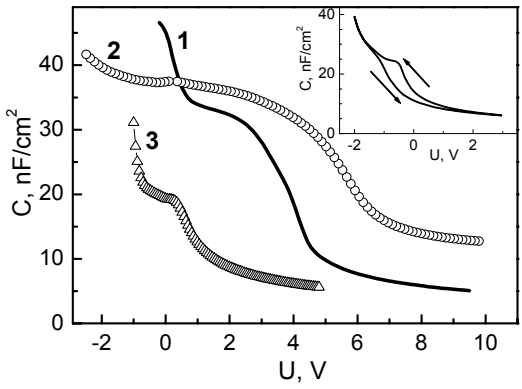

Figure 2

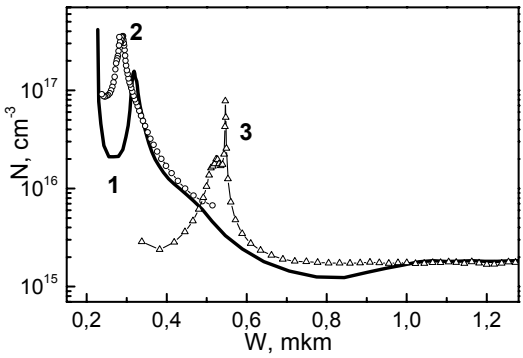

Figure 3

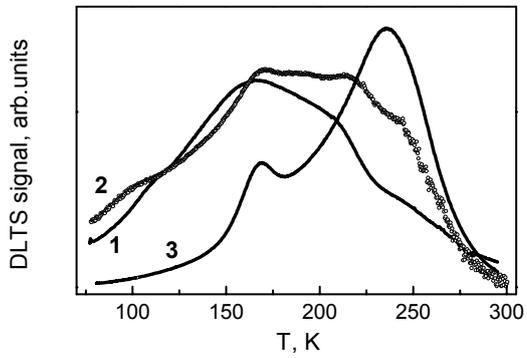

Figure 4